# Structural, chemical and optical characterizations of nanocrystallized AlN:Er thin films prepared by r. f. magnetron sputtering

V. Brien[a], P. Miska[b], H. Rinnert[b], D. Genève[a], P. Pigeat[a]

a. CNRS, Laboratoire de Physique des Milieux Ionisés et Applications, UMRCNRS 7040, Université de Nancy, Faculté des Sciences et Techniques, Boulevard des Aiguillettes, B. P. 239, F-54506 Vandoeuvre-lès-Nancy Cedex, France

b. Université de Nancy, Laboratoire de Physique des Matériaux, UMRCNRS 7556, Faculté des Sciences et Techniques, Boulevard des Aiguillettes, B. P. 239, F-54506 Vandoeuvre-lès-Nancy Cedex, France

For correspondence: Valerie. Brien@lpmi. uhp-nancy. fr, Fax: 00. 33. 3. 83. 68. 49. 33

**Abstract**

Nanocrystalline n-AlN:Er thin films were deposited on (001) Silicon substrates by r. f. magnetron sputtering at room temperature to study the correlation between 1.54 μm IR photoluminescence (PL) intensity, AlN crystalline structure and Er concentration rate. This study first presents how Energy-Dispersive Spectroscopy of X-rays (EDSX) Er Cliff Lorimer sensitivity factor $\alpha = 5$ is obtained by combining EDSX and electron probe micro analysis (EPMA) results on reference samples. It secondly presents the relative PL intensities of nanocrystallized samples prepared with identical sputtering parameters as a function of the Er concentration. The structure of crystallites in AlN films is observed by transmission electron microscopy.

**Introduction**

Rare earth (RE)-doped III-nitride semiconductor thin films are of considerable interest due to their potential applications in light-emitting diode and electroluminescence devices fabrication [1-5]. From studies made on Er insertion in semiconducting matrixes as $SiO_2$ (band gap ≈ 9 eV), it is known that photoluminescence (PL) efficiency of Er (intra-4f emission near 1.54 µm) can be enhanced by the presence of dispersed Si nanocrystallites inside $SiO_2$ material [6-9]. The morphology of the doped film and the localization of the dopant specie of Er doped wide and direct band gap materials such as AlN (6.2 eV) are also known to play a role in the efficiency of PL, but the mechanisms are not fully understood yet [2, 5, 6, 11-13].

To further characterize the crossed influence of nanostructure (size, shape of grains,…) and Erbium doping on PL emission, the authors prepared Er doped nanocolumnar and nanostructured AlN thin films by r. f. magnetron sputtering. In this work, one presents only the characterization of the nanocolumnar films prepared with different Erbium contents. The crystalline structure is observed by transmission electron microscopy (TEM) and Er concentration is measured by electron dispersive spectroscopy of X-rays (EDSX). To obtain precise and reproducible measurements of Er concentration in AlN films and in particular to assess the EDSX sensitivity coefficient of Er in AlN matrix, combinations of electron probe micro analysis (EPMA) and Auger electron spectroscopy (AES) analyses were done. Then, PL emission near 1.54 µm was observed and recorded on different nanocolumnar crystallized AlN:Er samples.

1. **Sample preparation**

AlN:Er depositions were carried out at room temperature on Si (001) substrates by r. f. magnetron sputtering. The background pressure of the UHV deposition chamber, controlled by mass spectroscopy, was $10^{-6}$ Pa. The insertion of Er in samples was obtained by simply laying regularly small pieces of Er (purity of 99.9%) on the Al (purity of 99.99 %) target.

The diameter and the thickness of the target are 60 mm and 3 mm large respectively and the target-sample distance is 150 mm. In this study, bias voltage is set to 0 Volt. The target was systematically sputter cleaned for 15 min using an Ar plasma (cleaning conditions: P = 0.5 Pa, W = 300 W). A gas mixture of Ar and $N_2$ of high purity (99.999 %) was used for sputtering. The $N_2/(Ar+N_2)$ percentage in the gas mixture was set to 50 %. During the treatment, a controlled pumping valve and mass-flow controllers (2.5 sccm $N_2$ and 2.5 sccm Ar) were used to keep the total sputtering pressure P constant equal to 1 Pa measured using a MKS Baratron gauge. The temperature of the substrate was measured with a thermocouple. It was found that the temperature changes stayed below 50°C. It was concluded that the heating is only due to the plasma presence. The reactor is equipped with an interferential optical reflectometer for real-time control of the thickness and the growth rate of the deposited layer.

A batch of five samples was prepared with no return to the atmosphere. The five depositions were made consecutively with identical process parameters (sputtering power = 50 W, plasma working pressure = 0.5 Pa). Progressive decreasing rates of Er are expected in films from one film to another due to progressive Er pieces abrasion.

3. Results

3. 1. Chemical analysis

To our knowledge, except some publications mentioning the use of an absolute analysis technique (Rutherford Back Scatterring) [2, 14], the other works quote analyses by EDSX, EPMA (Electron Probe Micro-Analysis), XPS (X-ray Photoelectron Spectroscopy) or SIMS (Secondary Ion Mass Spectrometry) with no detail on the calibration of the technique. The point is however of high relevance. As a matter of fact, the amount of luminescent elements in the matrices has important consequences on PL intensity [6-7, 15-16]. To characterize the PL emission as a function of the erbium content, preliminary calibration of chemical analyses was done.

The EDSX analysis was performed by means of a PGT (Princeton Gamma-Tech) spectrometer mounted on a CM20 Philips microscope and equipped with an ultra thin window X-Ray detector. The analyses were carried out in nanoprobe mode with a diameter of the probe of 10 nm. The Cliff Lorimer coefficient was determined using the stoichiometric erbium oxide compound $Er_2O_3$ as a reference sample: a value of 5 was found (coefficient of Si is set to 1). All the sensitivity coefficients found in this study are compiled in table 1. The composition of the five samples was then measured and values are compiled in table 2. As expected, the content in erbium of the samples progressively decreases from the first sample to the last one, because of the depletion of the target in erbium.

To check the measurements, sample C was independently analysed by means of EPMA. The equipment is a 5 spectrometers equipped CAMECA SX100 apparatus. The analysis was performed using an acceleration voltage of 5 kV and a probe current of 40 nA. The Al, O, Si and Er elements were analysed, N content was obtained by completion to 100 %. Calibration was achieved using pure silicon (Si $K_\alpha$), alumina (Al $K_\alpha$ and O $K_\alpha$) and ErNiSi (Er $M_\alpha$) as a reference sample. A value of 0.4 atomic % was obtained. The detection threshold of erbium with this technique under these conditions is 0.06 atomic %.

Auger electron spectroscopy was also performed. AES Concentration profiles were performed on a Microlab VG 310D using an Ar etching gun VG microprobe EX05. Electrons energy of 10 kV, ions energy of 3 kV, emission intensity of 2.5 mA and sample current of 400 nA were used. To calibrate the analysis, the erbium oxide ($Er_2O_3$) sample was also used and allowed to calculate the Auger MNN transition coefficients of erbium (M, N stand for the quantum states shells): values are given in table 2 [17]. Unfortunately, in the concentration range of the analysed samples (samples C and the richest one in erbium A were analysed), no significant signal could be recorded; the detection is not sufficient.

To locate possible erbium segregation (particles size above 10 nm) inside the samples, EDSX mapping was done on cross sections of the samples: it was found that the density of X-rays displays an homogeneous distribution informing that erbium is homogeneously distributed throughout the films (size of probe of 10 nm).

**3. 2. Morphology and optical characterization of the films**

Cross-sectional TEM images of the films (Fig. 1) show that the AlN:Er films prepared at various Er contents are all nanocolumnar. After a known to be amorphous first adaptation layer, as when undoped AlN is grown in the same manner in the same conditions, the columns have developed following a classical Van der Drift mechanism exactly [18,19]. The average width of the columns is 15-30 nm. One notes that such grain sizes are smaller than the ones in the classical columnar films optimized for SAW (Surface Acoustic Wave) applications [20]. The contrast of the images also shows that the films contain a lot of defects: about here and there one can see parallel fringes revealing the presence of planar defects (cf. arrows in Fig. 1 a and b). Magnification of such defects was placed in Fig. 1g and h.

The optical evaluation of the five AlN:Er films of this study was characterized using photoluminescence spectra at room temperature. The PL was analyzed in the range 1400-1800 nm by a monochromator equipped with a 600 grooves/mm grating and by a photomultiplier tube cooled at 190 K. The 325 nm excitation was obtained from a He–Cd laser. The response of the detection system was precisely calibrated with a tungsten wire calibration source. The PL spectra are given in Fig. 2. The peak corresponding to the erbium emission at 1.54 µm is clearly visible for low erbium concentrations. No peak can be seen for an erbium concentration of 2.6 at %.

The evolution of the integrated intensity of the PL spectra is reported in Fig. 3. It shows a decrease of the $Er^{3+}$ emission at 1.54 µm as erbium concentration increases. This evolution is well known in Si and $SiO_2$ systems, especially in $SiO_2$ matrices doped with silicon nanocrystals. It is called concentration quenching, corresponding to a drop of the 1.54 $Er^{3+}$ PL intensity due to an important erbium concentration [6, 8-9]. It is known, in other respects, that AlN even though it is nanocolumnar, does not exhibit any PL signal (0 % $Er^{3+}$ content). This leads to the logical conclusion that there must be a maximal value of the PL intensity between the points 0 % and 0.3 %. From this maximum, an optimal $Er^{3+}$ concentration can be deduced. In our case, as the 1.54 PL signal only decreases, it can be assumed that the optimal concentration value has not been found. A study to find the concentration condition producing the most intense PL signal is under progress.

The evolution of the PL intensity versus the $Er^{3+}$ concentration has also been measured by [16] on similar samples prepared by the same technique. They observed that the PL increases as the $Er^{3+}$ content increases until it reaches a maximum for the highest concentration followed by a saturation mechanism. The range of composition was 0.6 – 3.5 % (measured by EPMA). As, on the same composition range, our results exhibit an opposite trend, one has to conclude that other parameters than the amount of $Er^{3+}$ are critical to explain the PL intensity. One could think that the annealing the samples of Oliveira have underwent, has changed the local environment of erbium, the morphology etc…

All this demonstrates that more studies have to be done on this system to specify, understand and decorrelate the respective roles of the erbium content and environment, the grain shape and size, the oxygen content and location and annealing of the films.

## 4. Conclusions

This work allowed establishing different sensitivity coefficient of erbium in AlN of two chemical analysis techniques (ESDX, AES).

The integrated photoluminescence intensity is presented as a function of erbium content in the films over the 0.3 – 2.6 atomic % range. This set of results shows the integrated PL intensity decreases with increasing erbium content suggesting the presence of an optimal value for an erbium content located between 0 and 0.3%.

**Acknowledgments**

The authors wish to thank J. Ghanbaja for performing the TEM and EDSX observations, and S. Mathieu for EPMA analyses.

List of Figures and tables captions

Fig. 1: Morphology of the films obtained by TEM (from a/ to f/ micrographs on the left are bright field images, micrographs on the right are dark field images). a/ and b/ Sample A, c/ and d/ Sample C, e/ and f/ Sample E, g/ and h/ Magnification of the planar defects, they exhibit a fringes contrast.

Fig. 2: Photoluminescence spectra of the Er:doped nanocrystallized deposited on Si substrates as a function of the different erbium contents. The PL spectra were obtained at room temperature with a 325 nm excitation. They are vertically shifted for clarity reasons.

Fig. 3: 1. 5 µm integrated PL intensity as a function of the Er content (measured by calibrated EDSX).

Table 1: Coefficients obtained by calibration of the EDSX and AES techniques

Table 2: EDSX measurements of erbium contents in the five samples prepared for this study

Table 1

| EDSX | AES | |
|---|---|---|
| Cliff Lorimer* coefficient of erbium | Energy of the Auger erbium MNN transitions | Associated coefficients |
| 5 | 163 eV | 0.20 |
| | 1057 eV | 0.12 |
| | 1225 eV | 0.12 |
| | 1393 eV | 0.10 |

* The one of Silicon was set to 1

Table 2

| Sample | A | B | C | D | E |
|---|---|---|---|---|---|
| Er Atomic % | 2.6 | 1.9 | 0.5 | 0.4 | 0.3 |

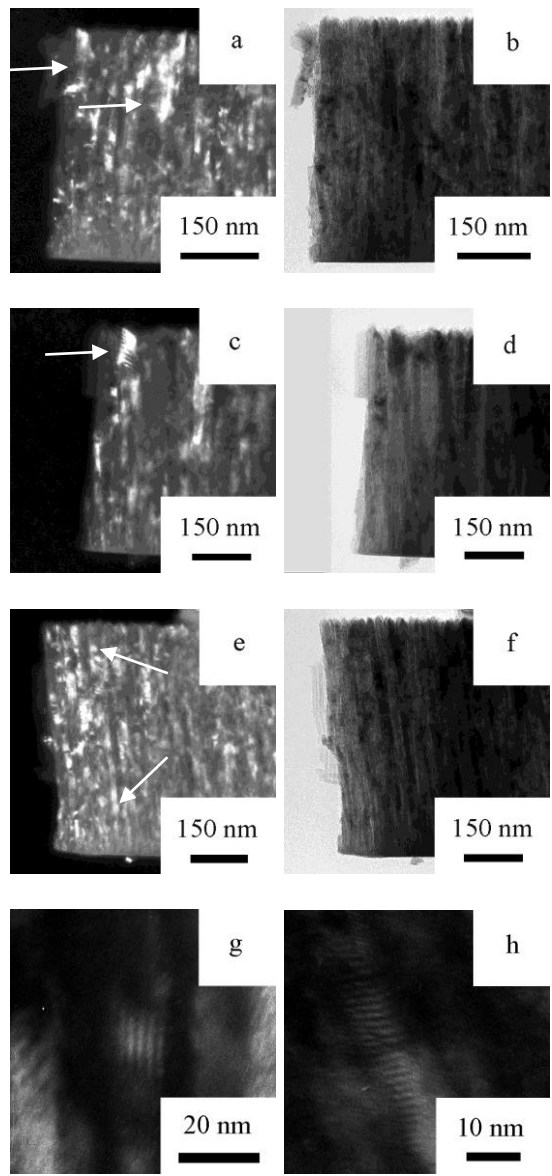

Figure 1

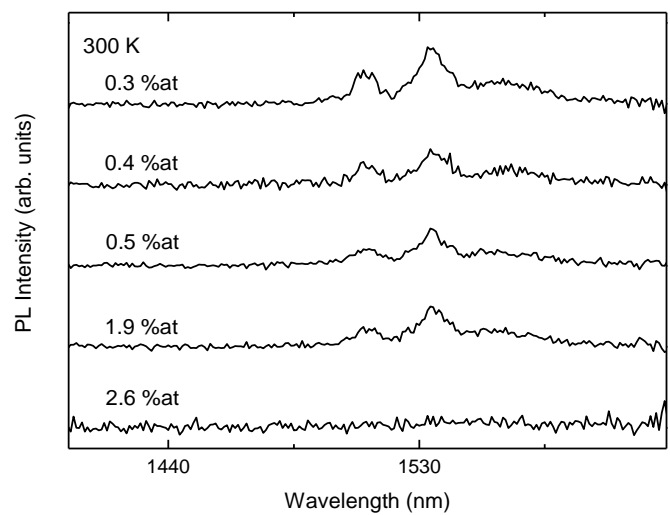

Figure 2

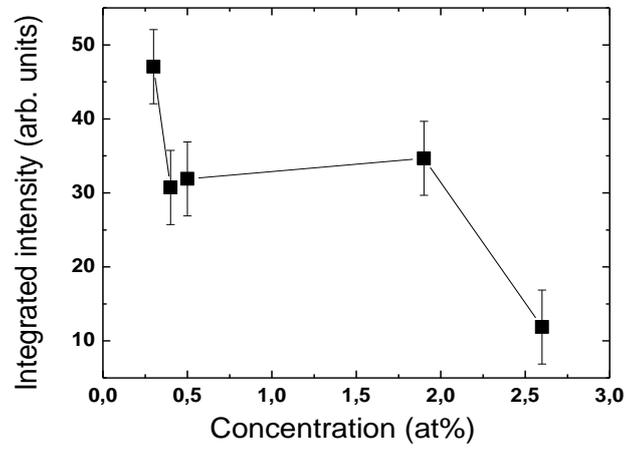

Figure 3